\documentclass[twocolumn,aps,showpacs]{revtex4}

\usepackage{epsfig}
\makeatletter

\newcommand{\boldsymbol}[1]{\mbox{\boldmath $#1$}}

\makeatother
\begin{document}

\title{Longitudinal spin waves in a dilute Bose gas}

\author{J. E. Williams\( ^{1} \), T. Nikuni\( ^{2} \) and Charles W. Clark\( ^{1} \)}

\affiliation{\( ^{1} \)National Institute
of Standards and Technology, 
Technology Administration,
U.S. Department of Commerce,
Gaithersburg, Maryland 20899-8410}

\affiliation{\( ^{2} \)Department of Physics, University of Toronto,
Toronto, Ontario, Canada M5S 1A7}

\date{\today}

\begin{abstract}
We present a kinetic theory for a dilute noncondensed Bose gas of
two-level atoms that predicts the transient spin segregation observed
in a recent experiment. The underlying mechanism driving spin currents
in the gas is due to a mean field effect arising from the quantum
interference between the direct and exchange scattering of atoms in
different spin states. We numerically solve the spin
Boltzmann equation, using a one dimensional model, and find excellent
agreement with experimental data.
\end{abstract}


\maketitle 

A recent experiment at JILA has displayed remarkable effects of spin
density fractionation in a trapped, ultra-cold gas of Rb atoms with no
Bose-Einstein condensate present~\cite{Lewandowski}. Under conditions
which we summarize in more detail below, sudden preparation of all
atoms in a coherent superposition of two spin states generates a spin
wave resulting in the observed spatial separation of the two
components. This occurs even though both the mean field and
differential Zeeman energy differences are almost a thousand times
smaller than the thermal energy $k_{\rm B}T$. In this paper, we show
that this astonishing departure from equilibrium results from quantum
interference between direct and exchange scattering of atoms in the
two spin states. A first-principles kinetic theory with no fitting
parameters gives excellent agreement with the experimental data, and
suggests possibilities for quantitative studies of quantum coherence
in noncondensed Bose-Einstein gases.

In the experiment of Lewandowski \emph{et al.}, a few million atoms of
$^{87}$Rb are confined in a magnetic cigar-shaped trap (\( \omega
_{z}/2\pi =7 \) Hz, \( \omega _{r}/2\pi =230 \) Hz). By applying
microwave and radio-frequency radiation, all atoms in the gas can be
uniformly prepared in an arbitrary superposition of the \(
|F=1,M_{F}=-1\rangle \equiv |1\rangle \) and \( |2,1\rangle \equiv
|2\rangle \) hyperfine states of the ground configuration.  The
frequency splitting, \( \Delta \equiv \omega _{1}-\omega _{2} \),
between the two states depends on the position, {\bf r} of the atom in
the trap: \( \Delta (\mathbf{r})=\Delta
_{\mathrm{BR}}(\mathbf{r})+\Delta _{\mathrm{MF}}(\mathbf{r})-\omega
_{\mathrm{hf}} \), where \( \omega _{\mathrm{hf}} \) is an overall
uniform frequency splitting (\( \omega _{\mathrm{hf}}/2\pi \sim 6.8 \)
GHz). The first term \( \Delta _{\mathrm{BR}}(\mathbf{r}) \) is due to
the differential Zeeman effect, predicted by the Breit-Rabi formula,
for atoms in a nonuniform magnetic field. The second term is due to
the mean field frequency shift proportional to the density of the gas,
which has a Gaussian profile in the harmonic trap. By applying two \(
\pi /2 \) pulses separated by a variable delay time, this local
frequency splitting is extracted from the Ramsey interference fringes
measured at different positions along the axial direction of the trap.

Lewandowski \emph{et al.} describe experiments in which the second \(
\pi /2 \) pulse is omitted and the time evolution of the density of
either state is observed after the initial \( \pi /2 \) pulse. The
following spectacular behavior is observed: the densities of the two
states segregate along the axial direction of the trap and then relax
to a completely overlapping stationary state after approximately \(
200 \) ms.  This fascinating behavior is found to depend crucially on
two different parameters: the density of the gas \( n \) and the
nonuniformity of the local frequency splitting \( \Delta (\mathbf{r})
\). No segregation is observed when the density is lowered below a
critical value, or \( \Delta (\mathbf{r}) \) is made approximately
uniform by adjusting the bias magnetic field. On the other hand, the
segregation effect becomes more dramatic as the density and the
inhomogeneity of the splitting are increased.

In this letter, we show that the transient spin segregation is
actually an overdamped spin wave arising solely from the mean field of
the gas, even if the interaction is spin independent. This effect
is well known from earlier work done on spin polarized hydrogen
gases~\cite{Leggett68,Lhuillier1,Levy84,Johnson84}, and is initiated
by the spatially varying local frequency splitting
$\Delta({\mathbf{r}})$. When $\Delta({\mathbf{r}})$ is uniform, the
spins of the atoms throughout the gas precess in exactly the same
fashion so that every forward scattering event can be understood in
terms of two identical atoms colliding, giving rise to the well-known
factor of $2$ from the direct and exchange terms in the Hartree-Fock
mean-field theory of a Bose gas. However, when $\Delta({\mathbf{r}})$
depends on position, two colliding atoms will have acquired different
spin states depending on their history in the trap. In this case, when
the direct and exchange forward scattering events are added, an
additional mean field term appears proportional to the local spin
$\vec{S}(\mathbf{r},t)$ of the gas, which accounts for the
constructive and destructive interference between the two scattering
paths. It is this term that gives rise to the spin wave. This mean
field effect occurs when the transverse spin (the internal coherence)
is long lived compared to thermal relaxation, emphasizing that the gas
of two-level atoms in the JILA experiment is quite different from an
incoherent binary mixture.  A solution of the collisionless Boltzmann
equation for the spin is already sufficient to predict spin
waves. However, because the JILA experiment is in an intermediate
regime approaching the hydrodynamic region, the spin current is
strongly damped due to collisions.

The Hamiltonian describing a single, trapped, two-level atom of mass
$m$ is:\begin{equation}
\label{H1}
\hat{H}=\left[-\frac{\hbar ^{2}}{2m}\nabla
^{2}+U_{\mathrm{ext}}(\mathbf{r}) \right]\hat{1}+\frac{\hbar
}{2}\vec{\Omega }(\mathbf{r})\cdot \vec{\hat{\tau }}.
\end{equation}
The first term in (\ref{H1}) is the center of mass Hamiltonian
containing the kinetic energy and the external parabolic trap \(
U_{\mathrm{ext}}(\mathbf{r})=m\omega _{z}^{2}\left[ \alpha
^{2}(x^{2}+y^{2})+z^{2}\right] /2 \), where \( \alpha =\omega
_{r}/\omega _{z} \). This part of the Hamiltonian is uncoupled from
the internal, pseudo-spin, degree of freedom, which is governed by the
second term: \( \vec{\Omega }(\mathbf{r})\cdot \vec{\hat{\tau
}}=\Omega _{x}(\mathbf{r})\hat{\tau }_{x}+\Omega
_{y}(\mathbf{r})\hat{\tau }_{y}+\Omega _{z}(\mathbf{r})\hat{\tau
}_{z}, \) where \( \hat{\tau }_{i} \) is a Pauli matrix.  In the
absence of an external coupling field, $\Omega_z=\Delta_{\mathrm{BR}}$
(we make the rotating wave approximation to eliminate the hyperfine
splitting $\omega_{\mathrm{hf}}$). We model binary interactions
between particles by a delta pseudo potential describing elastic, spin
preserving collisions, the strength of which depends on the hyperfine
states $V_{ij}(\mathbf{r},\mathbf{r}')= g_{ij}\delta
(\mathbf{r}-\mathbf{r}')$, where $g_{ij}=4\pi \hbar ^{2}a_{ij}/m$,
with $a_{ij}$ being the scattering length for collisions between atoms
of species $i$ and $j$.  For ${}^{87}$Rb, we take \( a_{11}=100.9a_{0}
\), \( a_{12}=98.2a_{0} \), \( a_{22}=95.6a_{0} \), where \( a_{0} \)
is the Bohr radius~\cite{Lewandowski}.

Several groups have previously worked out the fundamental kinetic
theory of a noncondensed dilute Bose gas with internal degrees of
freedom, to describe spin waves in spin-polarized atomic
hydrogen~\cite{Lhuillier1,Lhuiller2,Lhuillier3,Bouchaud85,Jeon,Ruckenstein89,Smith,NW}.
Using a semiclassical approximation to describe atomic motion in terms
of a phase space distribution function, we obtain coupled Boltzmann
equations for the distribution functions of atomic density, \(
f(\mathbf{r},\mathbf{p},t) \), and spin density, \( \vec{\sigma
}(\mathbf{r},\mathbf{p},t) \):
\begin{equation}
\label{kineticfcm}
\frac{\partial f}{\partial t}+\frac{\mathbf{p}}{m}\cdot \boldsymbol \nabla f-\boldsymbol \nabla U_{n}\cdot \boldsymbol \nabla _{p}f-\frac{\hbar }{2}\boldsymbol \nabla \Omega _{n\alpha }\cdot \boldsymbol \nabla _{p}\sigma _{\alpha }=\left. \frac{\partial f}{\partial t}\right| _{\mathrm{coll}} ,
\end{equation}
\begin{equation}
\label{kineticfs}
\frac{\partial \vec{\sigma }}{\partial t}+\frac{\mathbf{p}}{m}\cdot
\boldsymbol \nabla \vec{\sigma }-\boldsymbol \nabla U_{n}\cdot
\boldsymbol \nabla _{p}\vec{\sigma }-\frac{\hbar }{2}\boldsymbol
\nabla \vec{\Omega }_{n}\cdot \boldsymbol \nabla _{p}f-\vec{\Omega
}_{n}\times \vec{\sigma }=\left. \frac{\partial \vec{\sigma
}}{\partial t}\right| _{\mathrm{coll}} .
\end{equation}
Eq. (\ref{kineticfcm}) has an implicit sum over the repeated index
$\alpha$.  The total density and spin density are obtained from the
distribution functions as \( n(\mathbf{r},t)\equiv n_1({\bf
r},t)+n_2({\bf r},t)= \int d\mathbf{p}f(\mathbf{r},\mathbf{p},t)/(2\pi
\hbar )^{3} \) and \( \vec{S}(\mathbf{r},t)= \int
d\mathbf{p}\vec{\sigma}(\mathbf{r},\mathbf{p},t)/(2\pi \hbar)^{3} \)
respectively. Here the longitudinal component of the spin represents
the relative density $S_z=n_1-n_2$ and the transverse components $S_x$
and $S_y$ describe the real and imaginary parts of the internal
coherence.  The center of mass effective potential is \(
U_{n}(\mathbf{r},t)=U_{\mathrm{ext}}(\mathbf{r})+g_{11}n_{1}+g_{22}n_{2}+g_{12}(n_{1}+n_{2})/2. \)
The modified coupling field including mean-field effects
is\begin{equation}
\label{Omegan}
\vec{\Omega }_{n}(\mathbf{r},t)=\vec{\Omega }_{n}'(\mathbf{r},t)+{g_{12}\over \hbar }\vec{S}(\mathbf{r},t),
\end{equation}
 where \( \vec{\Omega }_{n}'(\mathbf{r},t)=\{\Omega
_{x}(\mathbf{r}),\, \Omega _{y}(\mathbf{r}),\, \Omega
_{z}(\mathbf{r})+\Delta _{\mathrm{MF}}(\mathbf{r},t)\} \), and
\begin{equation}
\label{Deltan}
\Delta_{\mathrm{MF}}(\mathbf{r},t)=
2\left[g_{11}n_{1}+g_{12}n_{2}-(g_{22}n_{2}+g_{12}n_{1})\right]/\hbar.
\end{equation}

The collision integral in
Eq.~(\ref{kineticfs}) is given by
\begin{eqnarray}
\left.\frac{\partial\vec{\sigma}}{\partial t}\right|_{\rm coll}
&=&\frac{\pi g_{12}^2}{\hbar} \int \frac{d{\bf p}_2}{(2\pi\hbar)^3}\int
\frac{d{\bf p}_3}{(2\pi\hbar)^3} \int d{\bf p}_4 \cr
&&\delta(\epsilon_p+\epsilon_{p_2}-\epsilon_{p_3}-\epsilon_{p_4})
\delta({\bf p}+{\bf p}_2-{\bf p}_3-{\bf p}_4) \cr
&&\times\{3f({\bf p}_3)\vec{\sigma}({\bf p}_4)
+\vec\sigma({\bf p}_3)f({\bf p}_4) \cr
&&-f({\bf p})\vec{\sigma}({\bf p}_2)
-3\vec{\sigma}({\bf p})f({\bf p}_2)\},
\label{collision}
\end{eqnarray}
where $\epsilon_p\equiv p^2/2m$. Here we neglect a principal value
contribution, which gives a second-order
correction to the free streaming evolution, and
we take all scattering lengths $a_{ij}$ to be equal
- a reasonable approximation for ${}^{87}$Rb. This approximation 
results in the conservation of spin density
during collisions, i.e. $\int d{\bf p} \partial\vec{\sigma}/
\partial t|_{\rm coll}=0$. When the small differences in scattering
lengths are accounted for, the transverse spin decays 
slowly. For ${}^{87}$Rb, this
contribution to the ``T2'' lifetime is of the order of
10 s~\cite{NW}.

Immediate insight can be gained if we solve for the time evolution 
of the spin density by integrating (\ref{kineticfs}) over momentum
\begin{equation}
\frac{\partial\vec{S}}{\partial t}+ \frac{1}{m}
\boldsymbol\nabla\cdot\vec{\bf J}=\vec\Omega_n'\times\vec{S}.
\label{eqfors}
\end{equation}
Since $\vec S\times \vec S=0$, the second term
$g_{12} \vec S / \hbar$ in (\ref{Omegan}) has no direct affect on the
spin density. We show below that this term instead sets up a spin
current $\vec{\bf J}$ in the gas that strongly affects $\vec S$. Also,
an interesting paradox has emerged concerning the factor of 2 in the
mean-field frequency shift. For an incoherent binary mixture of atoms
in either of the states $|1\rangle$ or $|2\rangle$
(i.e. $\sigma_x=\sigma_y=0$), it is straightforward to show from
(\ref{Omegan}) and (\ref{Deltan}) that the difference in chemicalpotentials due to interactions is $\mu_1-\mu_2=[2 (g_{11} n_1 - g_{22}
n_2) + g_{12} (n_2 - n_1)]$. There is a factor of 1 instead of 2 in
front of $g_{12}$ since the two atoms are distinguishable. However,
when the atoms are manipulated in a coherent fashion, from
(\ref{eqfors}) we see that the precession of the transverse spin is
given by $\Delta_{\mathrm{MF}}$, which is the mean-field frequency
shift measured by the Ramsey interference technique.

The hydrodynamic equation for the spin current $\vec{\bf J}$ is
obtained by assuming the following simple form for the spin
distribution function $\vec{\sigma}({\bf r},{\bf p},t)=f_0({\bf
r},{\bf p})[ \vec{M}({\bf r},t)+{\bf p}\cdot\vec{\bf v}({\bf
r},t)/k_\mathrm{B}T]$, where $\vec M({\mathbf{r}},t)\equiv \vec
S({\mathbf{r}},t)/n_0({\mathbf{r}})$ is the renormalized spin density.
Here, and in the rest of this letter, we assume the that the total
phase space density is stationary~\footnote{This assumption is
consistent with experimental observation~\cite{Lewandowski} and we
have checked it in our full numerical calculations using four coupled
equations for $f$ and $\vec{\sigma}$.} \(
f_{0}(\mathbf{r},\mathbf{p})=\eta _{0}\exp
\{-[p^{2}/2m+U_{\mathrm{ext}}(\mathbf{r})]/k_{\mathrm{B}}T\} \). The
constant \( \eta _{0}=\alpha ^{2}N(\hbar \omega
_{z}/k_{\mathrm{B}}T)^{3} \) is determined by the requirement \( \int
d\mathbf{p}\int d\mathbf{r}f_{0}/(2\pi \hbar )^{3}=N \), where \( N \)
is the total number of atoms. The equilibrium total density is
$n_0({\mathbf{r}})=\int d\mathbf{p} f_0({\mathbf{r}},{\mathbf{p}})/
(2\pi\hbar)^3$. Using this ansatz, the equation of motion for the spin
current $\vec{\bf J}\equiv \int d\mathbf{p} \mathbf{p}\vec \sigma /
(2\pi\hbar)^3$ is
\begin{equation}
\frac{\partial \vec{\bf J} } {\partial t}
-\vec{\Omega}_n\times\vec{\bf J} +k_{\rm B} T \boldsymbol\nabla\vec{S}
+\vec{S} \boldsymbol\nabla
U+\frac{\hbar}{2}n_0\boldsymbol\nabla\vec{\Omega}_n
=-\frac{\vec{\bf J}}{\tau_D}.
\label{eqforJ}
\end{equation}
Here, the diffusion relaxation time is~\cite{Smith,NW}
$\tau_D({\mathbf{r}})=[(32 a_{12}^2 n_0({\mathbf{r}})/3) \sqrt{\pi
k_{\mathrm{B}} T/m}]^{-1}$.

The spin segregation dynamics is described by the longitudinal spin
$S_z$. Taking the $z$ component of (\ref{eqfors}) and (\ref{eqforJ})
gives
\begin{equation}
\frac{\partial S_z}{\partial t}=-\frac{1}{m}\boldsymbol\nabla\cdot{\bf
J}_z,
\label{eqforsz}
\end{equation}
\begin{equation}
\frac{\partial {\bf J}_z}{\partial t}={\bf F}_z+\frac{g_{12}}{\hbar}
({\vec{S}}\times{\vec{\bf J}})_z-\frac{{\bf J}_z}{\tau_D},
\label{eqforJz}
\end{equation}
where ${\bf F}_z\equiv -k_{\rm B}T \boldsymbol\nabla
S_z-S_z\boldsymbol\nabla U -\hbar n_0\boldsymbol\nabla\Omega_{nz}/2$.
In the absence of a coupling field $\Omega_x=\Omega_y=0$, the
evolution of $S_z$ is entirely due to the spin current ${\bf J}_z$ and
thus the total $S_z$ is conserved ${\partial \over {\partial t}} \int
dz S_z = 0$.  The current ${\bf J}_z$ is driven by the first two terms
on the right-hand side of Eq.~(\ref{eqforJz}).  The first term is due
to the mechanical force ${\bf F}$ arising from the spatial gradient of
the local energy splitting.  The second term represents the spin
current driven by the dynamics of the $S_x$ and $S_y$ components
through the term $g_{12} \vec{S}/\hbar$.  The third term represents
the diffusion transport process that gives rise to the damping of the
current.  As already mentioned in this paper and also discussed in
Ref.~\cite{Lewandowski}, the magnitude of the mechanical force ${\bf
F}$ is negligibly small for driving the spin current ${\bf J}_z$ in
the experimental situation.  To highlight the effect of the term $g_{12}
\vec{S}/\hbar$, we take the time derivative of
Eq.~(\ref{eqforJz}), work to first
order in $g_{12}$, and neglect ${\bf F}_z$, the relaxation term,
and terms of second order in $\vec{\bf J}$.  This gives
\begin{equation}
\frac{\partial^2 {\bf J}_z}{\partial t^2}\simeq-\frac{g_{12}k_{\rm
B}T}{\hbar} (\vec{S}\times\boldsymbol\nabla \vec{S})_z=-\frac{g_{12}k_{\rm
B}T}{\hbar}S_{\perp}^2 \boldsymbol\nabla\phi,
\label{eqforJz2}
\end{equation}
where $S_{\perp}$ and $\phi$ are the amplitude and phase angle of the
transverse spin component $S_x+iS_y=S_{\perp}e^{i\phi}$.
Eq.~(\ref{eqforJz2}) explicitly shows that the spatial gradient of the
phase angle induces the spin current ${\bf J}_z$. In a short period of
time right after the $\pi/2$ pulse, it is reasonable to assume that
the transverse spin components are undergoing rotation with the local
Larmor frequency $\phi({\bf r},t)\approx\Omega_{nz}({\bf r})t$.  With
this simple approximation, the induced spin current for short times is
given by ${\bf J}_z\simeq-g_{12}k_{\rm B}TS_{\perp}^2 ({\bf
r})\boldsymbol\nabla\Omega_{nz}({\bf r})t^3/6\hbar$. Taking
$\Omega_{nz}=\Delta_{\mathrm{BR}}+\Delta_{\mathrm{MF}}\approx-m\omega_{\rm
diff}^2z^2/2\hbar$~\footnote{The parameter $\omega_\mathrm{diff}$ is
determined by fitting $\Delta_{\mathrm{BR}}+\Delta_{\mathrm{MF}}$ to a
parabola in the center of the trap.} and the initial condition
$S_{\perp}({\bf r})=n_0(0)e^{-U_{\rm ext}/k_\mathrm{B}T}$, we find
that the initial evolution of $S_z$ after the $\pi/2$ pulse is given
by.
\begin{equation}
S_z({\bf r},t)\simeq-\frac{g_{12}n_0^2({\bf r})k_{\rm
B}T}{24\hbar^2}\omega_{\rm diff}^2 \left(1-\frac{2m\omega_z^2}{k_{\rm
B}T}z^2\right)t^4.
\label{sz_initial}
\end{equation}
The above formula predicts that $S_z$ has nodes at $z=\pm \sqrt{k_{\rm
B}T/2m\omega_z^2}$, which we have verified in the numerical
calculation below.  Eq.~(\ref{sz_initial}) identifies the characteristic
timescale $t_{\rm spin}=[24\hbar^2/g_{12}n_0(0)k_{\rm B}T\omega_{\rm
diff}^2]^{1/4}$ needed for the system to build up the $S_z$ component.
For the JILA parameters, $t_{\rm spin}\approx 30$ ms, which is
consistent with the delay time seen in experiments before the
segregation begins.

To complement our simplified analysis, we also numerically solve the
Boltzmann equation (\ref{kineticfs}).  Motivated by the 
observation that spin
segregation occurs only along the axial direction~\cite{Lewandowski}, 
we construct a one
dimensional model of the system by making the ansatz \(
\vec\sigma(\mathbf{r},\mathbf{p},t)=\vec\sigma(z,p,t)h_{0}(x,y,p_{x},p_{y})
\) and then averaging over $x$ and $y$. Here we take the
static profile in the radial direction to be of Gaussian form \(
h_{0}=\exp \{-[(p_{x}^{2}+p_{y}^{2})/2m+m\omega
_{r}^{2}(x^{2}+y^{2})/2]/k_{\mathrm{B}}T\} \).  We substitute this
ansatz into (\ref{kineticfs}) and integrate over the radial phase
space variables, which gives the following one--dimensional model
Boltzmann equation
\begin{equation}
\label{kineticfs1D}
\frac{\partial \vec{\sigma }}{\partial t}+\frac{p}{m}\frac{\partial
\vec{\sigma }}{\partial z}-\frac{\partial U_{\mathrm{ext}}}{\partial
z}\frac{\partial \vec{\sigma }}{\partial p}-\vec{\Omega }_{n}\times
\vec{\sigma }=\left. \frac{\partial \vec{\sigma }}{\partial t}\right|
_{1D}.
\end{equation}
Here we have made the approximation \( U_{n}=U_{\mathrm{ext}} \) and
we have dropped the fourth term in (\ref{kineticfs}) coupling the
center of mass and spin. The collision integral in one dimension
involves a phase space average in the radial direction \(
\left. \partial \vec{\sigma }/\partial t\right| _{1D}\equiv \int
_{xy}\left. \partial \vec{\sigma }/\partial t\right|
_{\mathrm{coll}}/\int _{xy}h_{0}, \) where we have introduced the
notation \( \int _{xy}\cdots \equiv \int dx\int dy\int dp_{x}\int
dp_{y}\cdots /(2\pi \hbar )^{2} \).  The radial averaging introduces a
scaling factor in the mean field terms, so that \( g_{ij}\rightarrow
g_{ij}'= g_{ij}/(2\lambda_{\rm dB}^2) \), where $\lambda_{\rm dB}$
is the thermal de Broglie wavelength. $g_{ij}'$ has the correct units
of energy times distance required in our one dimensional model.

Although the direct numerical simulation using the full expression for
the one dimensional collision integral derived from
Eq.~(\ref{collision}) is technically feasible, we introduce a simple
model for the relaxation
\begin{equation}
\label{rel_approx}
\left.\frac{\partial \vec{\sigma}}{\partial t}\right|_{1D}
=-\frac{1}{\tau_{\rm cl}(z)}[\vec{\sigma}(z,p,t)
-\vec{M}(z,t)f_0(z,p)],
\end{equation}
where $\tau_\mathrm{cl}(z)=[16 a_{12}^2 n_0(z) \sqrt{\pi
k_{\mathrm{B}} T/m}]^{-1}$ is the radially averaged mean collision
time, $f_0(z,p)\equiv f_0(\mathbf{r},\mathbf{p})/h_0$, and
$\vec{M}(z,t)=\vec{S}(z,t)/n_0(z)$.  Eq. (\ref{rel_approx}) contains
the essential properties of collisions: (i) it vanishes when the
distribution function has the local equilibrium form
$\vec{\sigma}({\bf r},{\bf p},t) \propto e^{-p^2/2mk_{\rm B}T}$, (ii)
it conserves the spin density. We note that the form
(\ref{rel_approx}) does not require the knowledge of the long-time
equilibrium solution for $\vec{S}({\bf r},t)$.

\begin{figure}
  \centerline{\epsfig{file=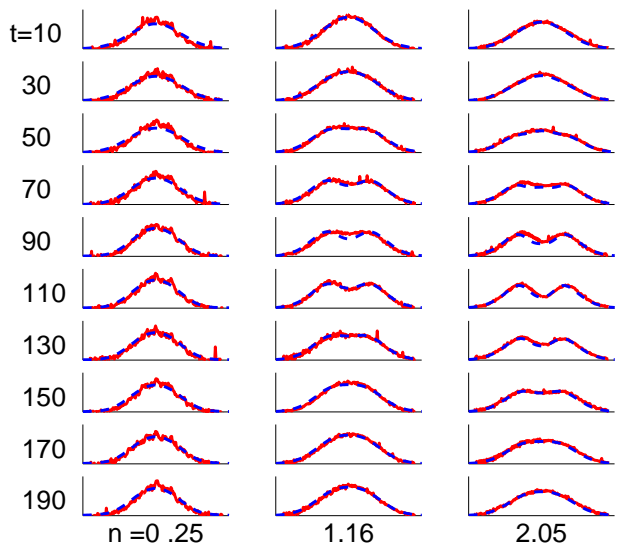,height=2.85in}}
\caption{Time sequence of $n_1(z,t)$. The red line is the
unsmoothed JILA data and the blue line is the numerical solution to
(\ref{kineticfs1D}). We have taken a temperature of $T=850$
nK~\cite{Lewandowski2}, and the peak total density $n$ listed under
each column is in units of $10^{13}$ cm${}^{-3}$. We approximate
$\Delta_{\mathrm{BR}}=m\omega_{\mathrm{BR}}^2 z^2/2\hbar$, and take
$\omega_{\mathrm{BR}}/\omega_z=0.09$ for each column in our
calculation. This corresponds to a value
$\omega_{\mathrm{diff}}/\omega_z=0.1$ for the first column.
The axial position $z$ is in the range $\pm 600 \, \mu$m.}
\end{figure}
We solved the one--dimensional spin kinetic equation
(\ref{kineticfs1D}) numerically using a finite difference scheme. For
the initial state of the spin, we take $\sigma_y(z,p,t=0)=f_0(z,p),
\sigma_x(z,p,t=0)=\sigma_z(z,p,t=0)=0$, corresponding to the state
immediately following the first $\pi/2$ pulse. Figure 1 shows the time
sequence of the density of the $|1\rangle$ state
$n_1(z,t)=[n_0(z)+S_z(z,t)]/2$, corresponding to Fig.3(c) columns
(v)-(vii) of Ref.~\cite{Lewandowski}. This shows that the spin
segregation vanishes when the density is lowered to $n\sim2\times
10^{12}$ cm${}^{-3}$. The red curve is the raw JILA data and the blue
dashed line is the theory. The agreement is striking. We also compare
our numerical results to Fig.3(c) columns (i)-(iv) of
Ref.~\cite{Lewandowski}, shown in Figure 2. In this case the density
is held fixed but the curvature of the energy splitting is varied in
each sequence. Note that when $\Delta_\mathrm{BR}(z)$ is chosen to
approximately cancel $\Delta_\mathrm{MF}$ in the second column, the
amplitude of the spin wave is essentially zero. For the larger values
of $\omega_\mathrm{diff}$ considered in columns three and four, the
agreement is only qualitative.
\begin{figure}
  \centerline{\epsfig{file=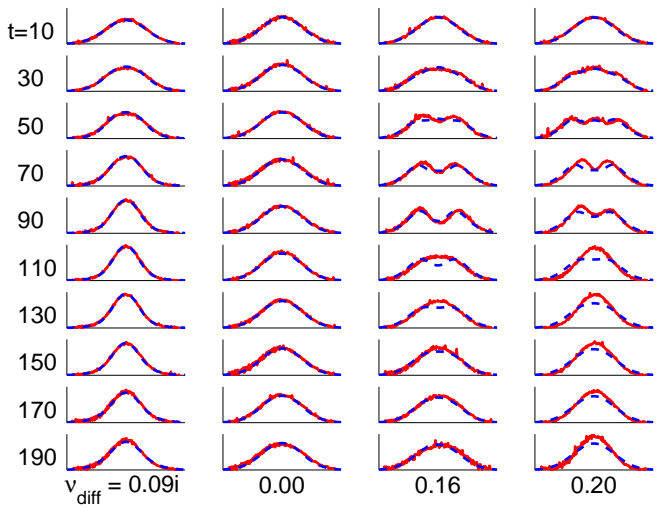,height=2.65in}}
\caption{Time sequence of $n_1(z,t)$. We have taken a temperature of
$T=850$ nK~\cite{Lewandowski2}, and the peak total density $n$ for
each column is $1.82 \times 10^{13}$ cm${}^{-3}$. Here,
$\nu_{\mathrm{diff}}=\omega_\mathrm{diff}/2\pi$ is in units of Hz.}
\end{figure}

\begin{figure}
  \centerline{\epsfig{file=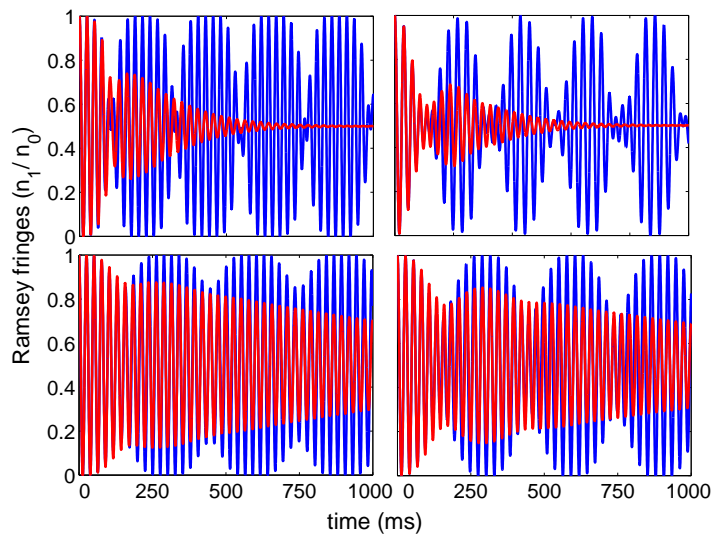,height=2.8in}}
\caption{Modulation of Ramsey fringes. The top row corresponds to the
third column of Figure 1, and in the bottom row we have set
$\Delta_{\mathrm{BR}}=0$ (which corresponds to sitting at the ``magic
spot'' bias field of 0.323 mT). The first column is taken at
$z=7\,\mu$m and the second column at $z=350\, \mu$m. The red and blue
lines compare the dynamics with and without collisional relaxation,
respectively. We have taken a detuning of $\delta/\omega_z = 5$. }
\end{figure}
We finally investigate the effect that the spin wave and relaxation
have on the Ramsey fringes. Ramsey fringes can be obtained from the
results of our calculation by simply rotating the Bloch vector
$\vec{S}(z,t)$ at each time by $90^\circ$ about the local oscillator
vector $\vec{\Omega}=\{\cos(\delta t),\sin(\delta t),0\}$, where
$\delta$ is the detuning between coupling field and the hyperfine
splitting. In the top row of Figure 3 we show the Ramsey fringes
corresponding to the third column of Figure 1, taken at the center
(left) and the edge (right) of the cloud.  We compare the Ramsey
fringes with (red) and without (blue) relaxation (i.e.
$\left. \partial \vec{\sigma }/\partial t\right| _{1D}=0$).  We find
that the Ramsey fringes taken at different positions across the cloud
are all modulated at the period of the spin wave, and that the fringe
visibility decays when the effect of collisions is included through
$\left. \partial \vec{\sigma }/\partial t\right| _{1D}$. At long times
the system relaxes to the state $\vec{S}=0$, that is, the gas evolves
to a completely overlapping binary mixture, which is reflected by the
vanishing of the fringe visibility. In the bottom row, the
differential Zeeman splitting is set to zero $\Delta_{\mathrm{BR}}=0$,
which reduces the curvature of the local energy splitting. This has
two main effects: the frequency of the spin wave is lowered and the
fringes are visible for a much longer time. In a calculation not shown
in Figure 3 for the case where the curvature is approximately zero, as
in column two of Figure 2, we find that the Ramsey fringes are not
modulated and do not decay, emphasizing that collisional relaxation
damps out spin currents, while conserving the spin. The trends shown
in the bottom row of Figure 3 agree qualitatively with
experiment~\cite{Lewandowski2}.

In summary, we have presented a simplified model for the spin density
and current that predicts longitudinal spin waves in the gas when the
energy splitting between hyperfine states in nonuniform. We have also
numerically solved a one dimensional model of the Boltzmann equation
that supports our simplified hydrodynamic model and gives excellent
agreement with the JILA experiment~\cite{Lewandowski}.

While this manuscript was being written, two independent
articles~\cite{Oktel,Fuchs} appeared on the lanl.arXiv.org e-print
archive that make predictions similar to ours.

We thank E. A. Cornell, H. J. Lewandowski, J. Roberts, and S. Rolston
for useful discussions.  We also thank H. J. Lewandowski for providing
us the experimental data used in Fig.~1.  T.N. acknowledges support
from JSPS.


\begin{thebibliography}{10}
\expandafter\ifx\csname bibnamefont\endcsname\relax
  \def\bibnamefont#1{#1}\fi
\expandafter\ifx\csname bibfnamefont\endcsname\relax
  \def\bibfnamefont#1{#1}\fi
\expandafter\ifx\csname url\endcsname\relax
  \def\url#1{\texttt{#1}}\fi
\expandafter\ifx\csname urlprefix\endcsname\relax\def\urlprefix{URL }\fi
\providecommand{\bibinfo}[2]{#2}
\providecommand{\eprint}[2][]{\url{#2}}

\bibitem{Lewandowski}
\bibinfo{author}{\bibfnamefont{H.~J.} \bibnamefont{Lewandowski}},
  \bibinfo{author}{\bibfnamefont{D.~M.} \bibnamefont{Harber}},
  \bibinfo{author}{\bibfnamefont{D.~L.} \bibnamefont{Whitaker}},
  \bibnamefont{and} \bibinfo{author}{\bibfnamefont{E.~A.}
  \bibnamefont{Cornell}}, \bibinfo{journal}{cond-mat/0109476} .

\bibitem{Leggett68}
\bibinfo{author}{\bibfnamefont{A.~J.} \bibnamefont{Leggett}} \bibnamefont{and}
  \bibinfo{author}{\bibfnamefont{M.~J.} \bibnamefont{Rice}},
  \bibinfo{journal}{Phys. Rev. Lett.} \textbf{\bibinfo{volume}{20}},
  \bibinfo{pages}{586} (\bibinfo{year}{1968}).

\bibitem{Lhuillier1}
\bibinfo{author}{\bibfnamefont{C.}~\bibnamefont{Lhuillier}} \bibnamefont{and}
  \bibinfo{author}{\bibfnamefont{F.}~\bibnamefont{Lalo\"e}},
  \bibinfo{journal}{J. Physique} \textbf{\bibinfo{volume}{43}},
  \bibinfo{pages}{197} (\bibinfo{year}{1982}).

\bibitem{Levy84}
\bibinfo{author}{\bibfnamefont{L.~P.} \bibnamefont{Levy}} \bibnamefont{and}
  \bibinfo{author}{\bibfnamefont{A.~E.} \bibnamefont{Ruckenstein}},
  \bibinfo{journal}{Phys. Rev. Lett.} \textbf{\bibinfo{volume}{52}},
  \bibinfo{pages}{1512} (\bibinfo{year}{1984}).

\bibitem{Johnson84}
\bibinfo{author}{\bibfnamefont{B.~R.} \bibnamefont{Johnson}},
  \bibinfo{author}{\bibfnamefont{J.~S.} \bibnamefont{Denker}},
  \bibinfo{author}{\bibfnamefont{N.}~\bibnamefont{Bigelow}},
  \bibinfo{author}{\bibfnamefont{L.~P.} \bibnamefont{Levy}},
  \bibinfo{author}{\bibfnamefont{J.~H.} \bibnamefont{Freed}}, \bibnamefont{and}
  \bibinfo{author}{\bibfnamefont{D.~M.} \bibnamefont{Lee}},
  \bibinfo{journal}{Phys. Rev. Lett.} \textbf{\bibinfo{volume}{52}},
  \bibinfo{pages}{1508} (\bibinfo{year}{1984}).

\bibitem{Lhuiller2}
\bibinfo{author}{\bibfnamefont{C.}~\bibnamefont{Lhuiller}} \bibnamefont{and}
  \bibinfo{author}{\bibfnamefont{F.}~\bibnamefont{Lalo\"e}},
  \bibinfo{journal}{J. Physique} \textbf{\bibinfo{volume}{43}},
  \bibinfo{pages}{225} (\bibinfo{year}{1982}).

\bibitem{Lhuillier3}
\bibinfo{author}{\bibfnamefont{C.}~\bibnamefont{Lhuillier}},
  \bibinfo{journal}{J. Physique} \textbf{\bibinfo{volume}{44}},
  \bibinfo{pages}{1} (\bibinfo{year}{1983}).

\bibitem{Bouchaud85}
\bibinfo{author}{\bibfnamefont{J.-P.} \bibnamefont{Bouchaud}} \bibnamefont{and}
  \bibinfo{author}{\bibfnamefont{C.}~\bibnamefont{Lhuillier}},
  \bibinfo{journal}{J. Physique} \textbf{\bibinfo{volume}{46}},
  \bibinfo{pages}{1101} (\bibinfo{year}{1985}).

\bibitem{Jeon}
\bibinfo{author}{\bibfnamefont{J.}~\bibnamefont{Jeon}} \bibnamefont{and}
  \bibinfo{author}{\bibfnamefont{W.}~\bibnamefont{Mullin}},
  \bibinfo{journal}{J. Physique} \textbf{\bibinfo{volume}{49}},
  \bibinfo{pages}{1691} (\bibinfo{year}{1988}).

\bibitem{Ruckenstein89}
\bibinfo{author}{\bibfnamefont{A.~E.} \bibnamefont{Ruckenstein}}
  \bibnamefont{and} \bibinfo{author}{\bibfnamefont{L.~P.} \bibnamefont{Levy}},
  \bibinfo{journal}{Phys. Rev. B} \textbf{\bibinfo{volume}{39}},
  \bibinfo{pages}{183} (\bibinfo{year}{89}).

\bibitem{Smith}
\bibinfo{author}{\bibfnamefont{H.}~\bibnamefont{Smith}} \bibnamefont{and}
  \bibinfo{author}{\bibfnamefont{H.~H.} \bibnamefont{Jensen}},
  \emph{\bibinfo{title}{Transport Phenomena}} (\bibinfo{publisher}{Clarendon
  Press}, \bibinfo{address}{Oxford}, \bibinfo{year}{1989}).

\bibitem{NW}
\bibinfo{author}{\bibfnamefont{T.}~\bibnamefont{Nikuni}} \bibnamefont{and}
  \bibinfo{author}{\bibfnamefont{J.~E.} \bibnamefont{Williams}},
  \bibinfo{note}{(unpublished)}.

\bibitem{Lewandowski2}
\bibinfo{author}{\bibfnamefont{H.~J.} \bibnamefont{Lewandowski}},
  \bibinfo{journal}{Private communication} .

\bibitem{Oktel}
\bibinfo{author}{\bibfnamefont{M.}~\bibnamefont{\protect\"O. Oktel}}
  \bibnamefont{and} \bibinfo{author}{\bibfnamefont{L.~S.}
  \bibnamefont{Levitov}}, \bibinfo{journal}{cond-mat/011119} .

\bibitem{Fuchs}
\bibinfo{author}{\bibfnamefont{J.~N.} \bibnamefont{Fuchs}},
  \bibinfo{author}{\bibfnamefont{D.~M.} \bibnamefont{Gangardt}},
  \bibnamefont{and} \bibinfo{author}{\bibfnamefont{F.}~\bibnamefont{Lalo\"e}},
  \bibinfo{journal}{cond-mat/0112228} .

\end{thebibliography}
\end{document}